\documentclass[preprint,showpacs,preprintnumbers,amsmath,amssymb,showkeys]{revtex4}

\usepackage{graphicx}
\usepackage{dcolumn}
\usepackage{bm}

\begin{document}

\preprint{}

\title{Focusing and phase compensation of paraxial beams by a left-handed material slab}

\author{Hailu Luo$^{1}$}\thanks{Author to whom correspondence should be addressed.
E-mail: hailuluo@163.com (H. Luo)}
\author{Wei Hu$^{2}$}
\author{Zhongzhou Ren$^{1}$}
\author{Weixing Shu$^{1}$}
\author{Fei Li$^{1}$}

\affiliation{$^{1}$ Department of Physics,
Nanjing University, Nanjing 210008, China\\
$^{2}$Laboratory of Light Transmission Optics, South China Normal
University, Guangzhou 510630, China}
\date{\today}

\begin{abstract}
On the basis of angular spectrum representation, a formalism
describing paraxial  beams propagating through an isotropic
left-handed material (LHM) slab is presented. The treatment allows
us to introduce the ideas of beam focusing and phase compensation
by LHM slab. Because of the negative refractive index of LHM slab,
the inverse Gouy phase shift and the negative Rayleigh length of
paraxial Gaussian beam are proposed. It is shown that the phase
difference caused by the Gouy phase shift in right-handed material
(RHM) can be compensated by that caused by the inverse Gouy phase
shift in LHM.  If certain matching conditions are satisfied, the
intensity and phase distributions at object plane can be
completely reconstructed at the image plane.
\end{abstract}

\maketitle

\section{Introduction}\label{Introduction}
In the late 1960s, Veselago firstly introduced the concept of
left-handed material (LHM) in which both the permittivity
$\varepsilon$ and the permeability $\mu$  are
negative~\cite{Veselago1968}. Veselago predicted that
electromagnetic waves incident on a planar interface between a
right-handed material (RHM) and a LHM will undergo negative
refraction. Theoretically, a LHM planar slab can act as a lens and
focus waves from a point source. Experimentally, the negative
refraction has been  observed by using periodic wires and rings
structure~\cite{Smith2000,Shelby2001,Pacheco2002,Parazzoli2003,Houck2003}.
In the past few years, negative refractions in photonic
crystals~\cite{Notomi2000,Luo2002a,Luo2002b,Li2003} and
anisotropic
metamaterials~\cite{Lindell2001,Hu2002,Smith2003,Zhang2003,Luo2005}
have also been reported.

Recently, Pendry  extended Veslago's analysis and further
predicted that a LHM slab can amplify evanescent waves and thus
behaves like a perfect lens~\cite{Pendry2000}. It is well known
that in a conventional imaging system the evanescent waves are
drastically decayed before they reach the image plane. While in a
LHM slab system, both the phases of propagating waves and the
amplitudes of evanescent waves from a near-field object could be
restored at its image. Therefore, the spatial resolution of the
superlens can overcome the diffraction limit of conventional
imaging systems and reach the subwavelength scale. While great
research interests were initiated by the revolutionary
concept~\cite{Smith2004a,Smith2004b,Parazzoli2004,Dumelow2005},
hot debates were also
raised~\cite{Hooft2001,Williams2001,Ziolkowski2001,Valanju2002,Garcia2002a,Garcia2002b,Vesperinas2004}.

The main purpose of the present work is to investigate the
paraxial beams propagating through an isotropic LHM slab. Starting
from the representation of plane-wave angular spectrum, we derive
the propagation of paraxial beams in RHM and LHM. Our formalism
permits us to introduce ideas for beam focusing and phase
compensation of paraxial beams by using LHM slab. Because of the
negative refractive index, the inverse Gouy phase shift and
negative Rayleigh length in LHM slab are proposed. As an example,
we obtain the analytical description for a Gaussian beam
propagating through a LHM slab. We find that the phase difference
caused by the Gouy phase shift in RHM can be compensated by that
caused by the inverse Gouy phase shift in LHM.

\section{The Paraxial model of beam propagation }\label{II}
In this section, we present a brief derivation on paraxial model
in RHM and LHM. Following the standard procedure, we consider a
monochromatic electromagnetic field ${\bf E}({\bf r},t) = Re [{\bf
E}({\bf r})\exp(-i\omega t)]$ and ${\bf B}({\bf r},t) = Re [{\bf
B}({\bf r})\exp(-i\omega t)]$ of angular frequency $\omega$
propagating through an isotropic material. The field can be
described by Maxwell's equations~\cite{Born1997}
\begin{eqnarray}
\nabla\times {\bf E} &=& - \frac{\partial {\bf B}}{\partial t} ,\nonumber\\
\nabla\times {\bf H} &=&  \frac{\partial {\bf D}}{\partial t},\nonumber\\
 {\bf D} &=& \varepsilon{\bf E},\nonumber\\
{\bf B} &=& \mu{\bf H}.
\end{eqnarray}
One can easily find that the wave propagation is only permitted in
the medium with $\varepsilon, \mu>0$ or $\varepsilon,\mu<0$. In
the former case, ${\bf E}$, ${\bf H}$ and ${\bf k}$ form a
right-handed triplet, while in the latter case, ${\bf E}$, ${\bf
H}$ and ${\bf k}$ form a left-handed triplet. The previous Maxwell
equations can be combined straightforwardly to obtain the
well-known equation for the complex amplitude of the electric
field in RHM or LHM
\begin{equation}
\nabla^2\ {\bf E}-\nabla (\nabla\cdot {\bf E})+  k^2 {\bf E}
=0,\label{ca}
\end{equation}
where $k =n_{R,L}\omega/c$, $c$ is the speed of light in vacuum,
$n_R=\sqrt{\varepsilon_R\mu_R}$ and
$n_L=-\sqrt{\varepsilon_L\mu_L}$ are the refractive index of RHM
and LHM, respectively~\cite{Veselago1968}.

Equation~(\ref{ca}) can be conveniently solved by employing the
Fourier transformations, so the complex amplitude in RHM and LHM
can be conveniently expressed as
\begin{eqnarray}
 {\bf E}({ {\bf r}_\bot},z )=\int d^2 {\bf k}_\bot
\tilde{E}({\bf k}_{\bot})\exp [i{\bf k}_{\bot}\cdot{\bf
r}_{\perp}+i k_z z] .\label{nop}
\end{eqnarray}
Here ${\bf r}_\perp=x{\bf e}_x+y{\bf e}_y$, ${\bf k}_\perp
=k_x{\bf e}_x+k_y {\bf e}_y$, and $e_j$ is the unit vector in the
$j$-direction. Note that $k_z= \sigma \sqrt{n_{R, L}^2 k^2_0 -
k^2_\perp}$, $\sigma=1$ for RHM and $\sigma=-1$ for LHM. The
choice of sign ensures that power propagates away from the surface
to the $+z$ direction.  The field $\tilde{E}({\bf k}_{\bot})$ In
Eq.~(\ref{nop}) is related to the boundary distribution of the
electric field by means of the relation
\begin{eqnarray}
\tilde{{\bf E}}({\bf k}_{\bot})=\int d^2 {\bf r}_\bot {\bf E}({\bf
r}_{\bot},0)\exp [i{\bf k}_{\bot}\cdot{\bf r}_{\perp}] ,\label{as}
\end{eqnarray}
which is a standard two-dimensional Fourier
transform~\cite{Goodman1996}. In fact, after the electric field on
the plane $z=0$ is known, Eq.~(\ref{nop}) together with
Eq.~(\ref{as}) provides the expression of the field in the space
$z > 0$.

From a mathematical point of view, the approximate paraxial
expression for the field can be obtained by the expansion of the
square root of $k_z$ to the first order in $|{\bf
k}_\bot|/k$~\cite{Lax1975,Ciattoni2000}, which yields
\begin{eqnarray}
 {\bf E}({ {\bf r}_\bot},z )=& &\exp(i n_{R,L} k_0 z) \int d^2 {\bf k}_\bot\nonumber\\
 & &\times\exp \bigg[i{\bf k}_{\perp}\cdot{\bf r}_{\perp}-\frac{i {\bf k}_\perp
z}{2n_{R,L} k_0}\bigg]\tilde{\bf E}({\bf k}_\perp).\label{field}
\end{eqnarray}
Since our attention will be focused on beam propagating  along the
$+z$ direction, we can write
\begin{equation}
{\bf E}({\bf r}_\perp,z)={\bf A}({\bf r}_\perp,z) \exp(i n_{R,L}
k_0 z),\label{peq}
\end{equation}
where the field $A( {\bf r}_\perp , z)$ is the slowly varying
envelope amplitude which satisfies the parabolic equation
\begin{equation}
\bigg[i\frac{\partial}{\partial z}+\frac{1}{2 n_{R,L}
k_0}\nabla_\perp^2 \bigg]{\bf
 A}({\bf r}_\perp,z)=0,\label{pe}
\end{equation}
where $\nabla_\perp=\partial_x {\bf e}_x+ \partial_y {\bf e}_y$.
From Eq.~(\ref{pe}) we can find that the field of paraxial beams
in LHM can be written in the similar way to that in RHM, while the
sign of the refractive index is reverse.

\section{The propagation of paraxial Gaussian beam}\label{SecIII}
\begin{figure}
\includegraphics[width=10cm]{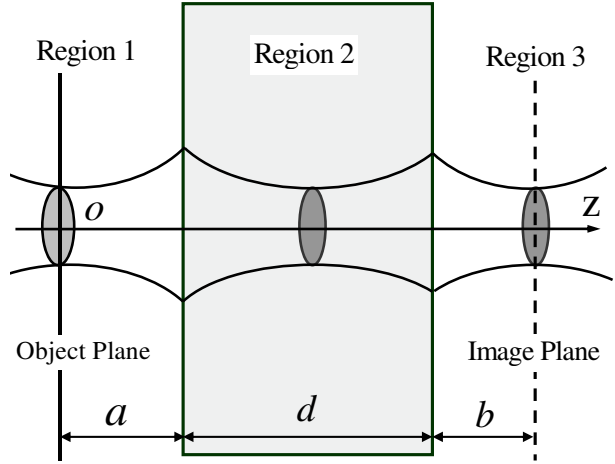}
\caption{\label{Fig1} The mechanisms for paraxial beams
propagating through an isotropic LHM slab. The LHM slab in region
$2$ is surrounded by the usual RHM in region $1$ and region $3$.
The solid line and the dashed line are the theoretical object and
image planes, respectively.}
\end{figure}

The previous section outlined the paraxial model for general laser
beams propagating in RHM and LHM. In this section we shall
investigate the analytical description for a beam with a boundary
Gaussian distribution. This example allows us to describe the new
features of beam propagation in LHM slab. As shown in
Fig.~\ref{Fig1}, the LHM slab in region $2$ is surrounded by the
usual RHM in region $1$ and region $3$. The beam will pass the
interfaces $z = a$ and $z = a+d$ before it reaches the image plane
$z=a+b+d$. To be uniform throughout the following analysis, we
introduce different coordinate transformations $z_i^\ast
(i=1,2,3)$ in the three regions, respectively.

First we want to explore the field in region $1$. Without any loss
of generality, we assume that the input waist locates at the
object plane $z=0$. The fundamental Gaussian spectrum distribution
can be written in the form
\begin{equation}
{\bf E}_1( {\bf
k}_\perp)=\frac{w_0}{\sqrt{2\pi}}\exp\bigg[-\frac{k_\perp^2
w_0^2}{4}\bigg],\label{s1}
\end{equation}
where $w_0$ is the spot size. By substituting Eq.~(\ref{s1}) into
Eq.~(\ref{field}), the field in the region $1$ can be written as
\begin{equation}
{\bf
 E}_{1}({\bf r}_\perp,z_1^\ast)=\frac{w_0}{w_{1}(z_1^\ast)}\exp\bigg[-\frac{r_\perp^2}
 {w_{1}^2(z_1^\ast)}+i\psi_{1}(r_\perp, z_1^\ast)\bigg],\label{g1}
\end{equation}
\begin{eqnarray}
\psi_{1} (r_\perp, z_{1}^\ast) &=&n_R k_0 z_1^\ast+ \frac{n_R k_0
r_\perp^2}{2 R_1(z_1^\ast)}-\arctan
\frac{z_1^\ast}{z_R},\label{p1}\\
w_{1}(z_{1}^\ast)&=&w_0 \sqrt{1+(z_{1}^\ast/z_R)^2},\\
R_{1}(z_{1}^\ast)&=&z_1^\ast+\frac{z_R^2}{z_1^\ast}.
\end{eqnarray}
Here $z_1^\ast=z$, $z_R= n_R k_0 w_0^2 /2$ is the Rayleigh length,
$w_{1}(z_{1}^\ast)$ is the beam size and $R_{1}(z_{1}^\ast)$ the
radius of curvature of the wave front. The first term and the
second term in Eq.~(\ref{p1}) are the plane wave phase and radial
phase, respectively. The third term in Eq.~(\ref{p1}) denotes the
Gouy phase is given by $\Phi_1=-\arctan (z_1^\ast/z_R)$.

We are now in a position to calculate the field in region $2$. In
fact, the field in the first boundary can be easily obtained from
Eq.~(\ref{g1}) by choosing $z=a$. Substituting the field into
Eq.~(\ref{as}), the angular spectrum distribution can be obtained
as
\begin{equation}
{\bf
 E}_{2}( {\bf k}_\perp)=\frac{w_0}{\sqrt{2\pi}}\exp\bigg[-\frac{k_\perp^2
(n_R k_0 w_0^2+2ia)}{4 n_R k_0}\bigg],\label{s2}
\end{equation}
For simplicity, we assume that the wave propagates through the
boundary without reflection. Substituting Eq.~(\ref{s2}) into
Eq.~(\ref{field}), the field in the LHM slab can be written as
\begin{equation}
{\bf
 E}_{2}({\bf r}_\perp,z_2^\ast)=\frac{w_0}{w_{2}(z_2^\ast)}\exp\bigg[-\frac{r_\perp^2}
 {w_{2}^2(z_2^\ast)}+i\psi_{2}(r_\perp, z_2^\ast)\bigg],\label{g2}
\end{equation}
\begin{eqnarray}
\psi_{2} (r_\perp, z_{2}^\ast) &=&n_L k_0 z_2^\ast+ \frac{n_L k_0
r_\perp^2}{2 R_2(z_2^\ast)}-\arctan
\frac{z_2^\ast}{z_L},\label{p2}\\
w_{2}(z_2^\ast)&=&w_0\sqrt{1+(z_2^\ast/z_L)^2},\label{w2}\\
R_{2}(z_2^\ast)&=&z_2^\ast+\frac{z_L^2}{z_2^\ast},\label{r2}
\end{eqnarray}
Here $z_2^\ast=z-(1-n_L/n_R)a$ and $z_L= n_L k_0 w_0^2 /2$ is the
Rayleigh length in LHM. The beam size $w_{2}(z_2^\ast)$ and the
radius of curvature $R_{2}(z_2^\ast)$ are given by Eq.~(\ref{w2})
and Eq.~(\ref{r2}), respectively. The Gouy phase shift in LHM is
given by $\Phi_2=-\arctan (z_2^\ast/z_L)$.

\begin{figure}
\includegraphics[width=10cm]{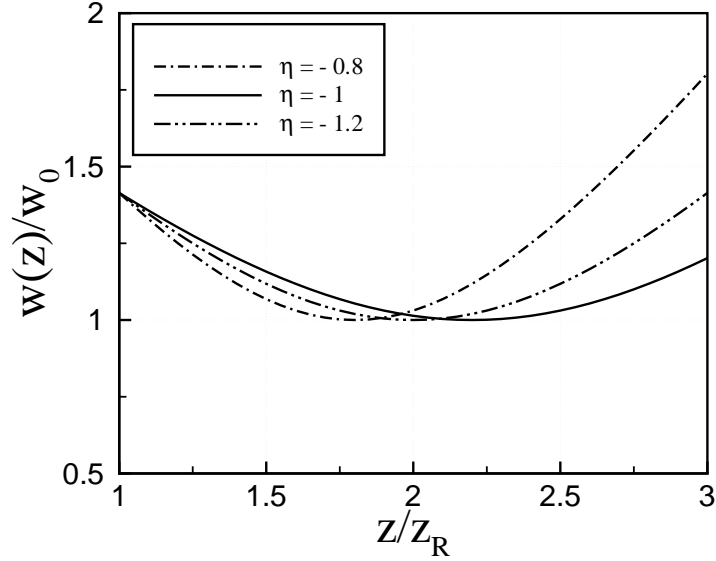}
\caption{\label{Fig2} Plot of the beam widths in LHM slab with
different refractive indices. The solid ($\eta=-1$),
dash-dot-dotted ($\eta=-1.2$), and dash-dotted ($\eta=-0.8$) lines
indicate the beam widths in LHM slab, where we assume $a=z_R$ and
$\eta=n_L/n_R$. One can find the beams can be focused by LHM slab
and the focusing waists remain $w_0$.}
\end{figure}

We note two interesting features of the paraxial field in LHM:
First, because of the negative index of refraction, the inverse
Gouy phase shift and negative Rayleigh length should be
introduced. Second, the beams can be focused by LHM slab and the
focusing waists remain $w_0$. For the purpose of illustration, the
relevant focusing feature is shown in Fig. \ref{Fig2}. We find
that the place of focusing waist could be different which depends
on the choice of the refractive index.

Finally we want to study the field in region $3$. The field in the
second boundary can be easily obtained from Eq.~(\ref{g2}) under
choosing $z=a+d$. Substituting the field into Eq.~(\ref{as}), the
angular spectrum distribution can be written as
\begin{equation}
{\bf
 E}_{3}( {\bf k}_\perp)=\frac{w_0}{\sqrt{2\pi}}\exp\bigg[-\frac{k_\perp^2
(n_R n_L k_0 w_0^2+2in_L a+2in_R d)}{4 n_R n_L
k_0}\bigg].\label{s3}
\end{equation}
Substituting Eq.~(\ref{s3}) into Eq.~(\ref{field}), the field in
the LHM slab is given by
\begin{equation}
{\bf
 E}_{3}({\bf r}_\perp,z_3^\ast)=\frac{w_0}{w_{3}(z_3^\ast)}\exp\bigg[-\frac{r_\perp^2}
 {w_{3}^2(z_3^\ast)}+i\psi_{3}(r_\perp, z_3^\ast)\bigg],\label{g3}
\end{equation}
\begin{eqnarray}
\psi_{3} (r_\perp, z_{3}^\ast) &=&n_R k_0 z_3^\ast+ \frac{n_R k_0
r_\perp^2}{2 R_3(z_3^\ast)}-\arctan
\frac{z_3^\ast}{z_R},\label{p3}\\
w_{3}(z_{3}^\ast)&=&w_0 \sqrt{1+(z_3^\ast/z_R)^2},\label{w3}\\
R_{3}(z_{3}^\ast)&=&z_3^\ast+\frac{z_R^2}{z_3^\ast}.\label{r3}
\end{eqnarray}
Here $z_3^\ast=z-(1- n_R/n_L)d$. The beam size $w_{3}(z_3^\ast)$
and the radius of curvature $R_{3}(z_3^\ast)$ are given by
Eq.~(\ref{w3}) and Eq.~(\ref{r3}), respectively. The Gouy phase
shift is given by $\Phi_3=-\arctan (z_3^\ast/z_R)$.

\begin{figure}
\includegraphics[width=10cm]{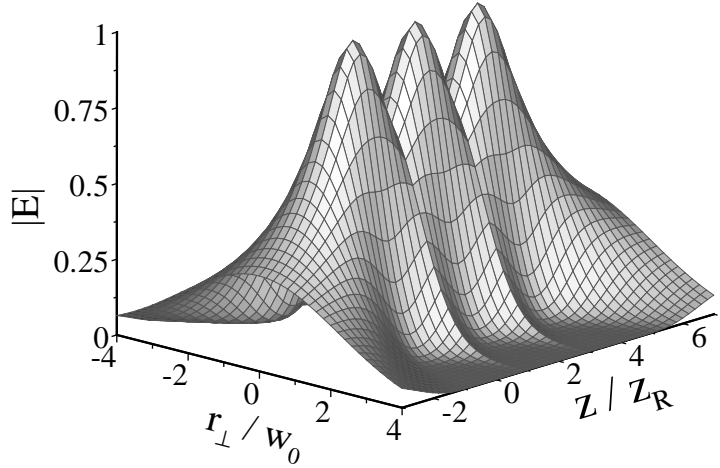}
\caption{\label{Fig3} Numerically computed spatial map of the
magnitude of the electric field for Gaussian beam propagating
through the LHM slab. We assume $a=z_R$, $d=2z_R$ and $n_L=-n_R$.
The first focusing waist in LHM slab locates at the plane
$z/z_R=2$, while the second focusing waist in region $3$ locates
at the plane $z/z_R=4$.}
\end{figure}

To this end, the fields are determined explicitly in the three
regions. Comparing Eq.~(\ref{g2}), Eq.~(\ref{g3}) with
Eq.~(\ref{g1}) show that the field distributions in region $2$ and
region $3$ still remain Gaussian. For the purpose of illustration,
the spatial map of the electric fields is plotted in Fig.
\ref{Fig3}. We can easily find that there exists an internal and
an external focus.

\section{Beam focusing and  phase compensation}\label{SecIV}
In this section we examine the matching conditions of beam
focusing and phase compensation. First we want to explore the
matching condition of focusing. We can easily find the place of
the focusing waist by choosing $z^\ast_i=0$. We assume the
incident beam waist locates at plane $z=0$. After setting
$z^\ast_2=0$ in Eq.~(\ref{w2}), we get the first focusing waist in
LHM slab locates at the plane $z=(1-n_L/n_R) a$. Then we
substitute $z^\ast_3=0$ into Eq.~(\ref{w2}), we can find the
second focusing waist in region $3$ locates at the plane $z=(1 -
n_R/n_L)d $. We take the image position $z=a+d+b$ to be the place
of the second focusing waist. Using this criterion, the matching
condition for focusing can be written as
\begin{equation}
n_L(a+b)+n_R d=0.\label{foc}
\end{equation}
In addition, the thickness of the LHM slab should satisfy the
relation $d>-a n_L /n_R$, otherwise there is neither an internal
nor an external focus.

\begin{figure}
\includegraphics[width=10cm]{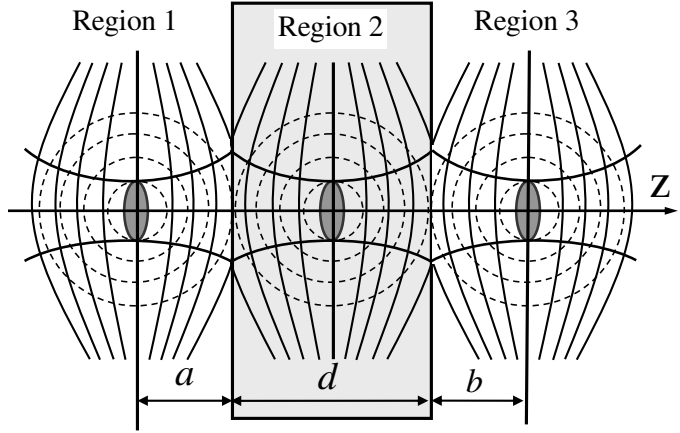}
\caption{\label{Fig4} The phase difference caused by the Gouy
phase shift in RHM can be compensated by that caused by the
inverse Gouy phase shift in LHM slab. The phase fronts of Gaussian
beam (solid lines) differ from those of a perfect spherical wave
(dashed lines).}
\end{figure}

Next we attempt to investigate the matching condition for phase
compensation. It is known that an electromagnetic beam propagating
through a focus experiences an additional $\pi$ phase shift with
respect to a plane wave. This phase anomaly was discovered by Gouy
in 1890 and has since been referred to as the Gouy phase
shift~\cite{Siegman1986,Feng2001}. It should be mentioned that
there exists an effect of accumulated Gouy phase shift  when a
beam passing through an  optical system with positive
index~\cite{Erden1997,Feng1999,Feng2000}. While in the LHM slab
system we expect that the phase difference caused by Gouy phase
shift can be compensated by that caused by the inverse Gouy shift
in LHM. In Fig.~\ref {Fig4}, we plot the distribution of phase
fronts in the three regions. The phase difference caused by the
Gouy phase shift in the three regions are given by
\begin{eqnarray}
\Delta \Phi_1&=&-\arctan\frac{a}{z_R},\nonumber\\
\Delta \Phi_2&=& -\arctan \frac{-n_L a}{n_R z_L}-\arctan
\frac{-n_L b}{n_R z_L},\nonumber\\
\Delta \Phi_3&=&-\arctan \frac{b}{z_R}.\label{gd}
\end{eqnarray}
The first and third Equations dictate the phase difference caused
by the Gouy shift in region $1$ and region $3$, respectively. The
second equation denotes the phase difference caused by the inverse
Gouy phase shift in LHM slab. Subsequent calculations of
Eq.~(\ref{gd}) show
\begin{equation}
-\arctan \frac{a}{z_R}-\arctan \frac{b}{z_R}+\bigg(-\arctan
\frac{-n_L a}{n_R z_L}-\arctan \frac{-n_L b}{n_R
z_L}\bigg)=0.\label{gouy}
\end{equation}
This means that the phase difference caused by the Gouy phase
shift in RHM can be compensated by that caused by the inverse Gouy
phase shift in LHM slab. Therefore the condition for phase
compensation can be simply written as
\begin{equation}
n_R k_0 a+n_R k_0 b+n_L k_0 d=0.\label{pha}
\end{equation}
The first two terms in Eq.~(\ref{pha}) are the phase deference
caused by the plane wave in RHM. The last term  in Eq.~(\ref{pha})
is the phase deference caused by the plane wave in LHM slab.

Finally we discuss the phase difference caused by the radial
phase. Following the method outlined by Dumelow {\it et
al.}~\cite{Dumelow2005}, we assume that the beam waist locates at
the object plane. Then we take the image position to be that where
the intensity is a maximum. Evidently, from
Eqs.~(\ref{g3})-(\ref{r3}) we can find that the intensity maximum
locates at the plane of beam waist. The phase difference between
the object plane and the image plane is independent of radial
position, since the phase fronts are flat there.

The new message in this paper is to prove that the intensity and
phase distributions at object plane can be completely
reconstructed at the image plane by LHM slab. Now an interesting
question naturally arises: whether the matching conditions of
focusing and the phase compensation can be satisfied
simultaneously. From Eqs.~(\ref{foc})-(\ref{pha}) one can easily
find that the refractive index, $a$, $b$ and $d$ should satisfy
the matching conditions:
\begin{equation}
n_L=-n_R, ~~~a+b=d. \label{del}
\end{equation}
Under the matching conditions, the reflected waves at the
interface between RHM and LHM are completely absent. Therefore the
intensity and phase distributions at the object plane can be
completely reconstructed at the image plane.

Note that the purpose of this paper is to examine beam focusing
and phase compensation in paraxial regime. The evanescent waves
which are claimed to provide the subwavelength imaging do not
correspond to the problem under study. The paraxial model only
deals with beams whose transverse dimension is much larger than a
wavelength. However, in the subwavelength focusing regime a
rigorous diffraction theory should be developed.

\section{Conclusions}
In conclusion, we have investigated the focusing and phase
compensation of paraxial beams by an isotropic LHM slab. We have
introduced the concepts of inverse Gouy phase shift and negative
Rayleigh length of paraxial beams in LHM. We have shown that the
phase difference caused by the Gouy phase shift in RHM can be
compensated by that caused by the inverse Gouy phase shift in LHM
slab. If certain matching conditions are satisfied, the intensity
and phase distributions at object plane can be completely
reconstructed at the image plane. We expect many potential devices
can be constructed based on the paraxial beam model discussed
above. They can, for example, be used to provide beam focusing,
phase compensation and image transfer.

\begin{acknowledgements}
H. Luo are sincerely grateful to Professors J. Ding, Q. Guo and L.
B. Hu for many fruitful discussions. This work was supported by
projects of the National Natural Science Foundation of China (No.
10125521, No. 10535010 and No. 60278013), the Fok Yin Tung High
Education Foundation of the Ministry of Education of China (No.
81058).
\end{acknowledgements}

\end{document}